\documentclass[twoside,12pt]{article}  %%% 这是 LaTeX 格式文件必须的指令，不可缺少
\usepackage{CJK}   %%%%% 使用中文、日文、韩文时调用此宏包
\usepackage{indentfirst}  %% 节标题后第一行段缩进
\usepackage{bm}    %%%%%  排印数学符号黑斜体时调用此宏包
\usepackage{graphicx}  %%%% 文中插入 eps 图形时调用此宏包，插图方法 1
\begin{document}    %% 文本文件开始，这是必须的指令

\begin{CJK*}{GBK}{song}  %% 开始进入中文环境

%-------------------  First Head  -----------------------------------------
\thispagestyle{empty} \vspace*{0.8cm}\hbox
to\textwidth{\vbox{\hfill\huge\sf \hfill}}
\par\noindent\rule[3mm]{\textwidth}{0.2pt}\hspace*{-\textwidth}\noindent
\rule[2.5mm]{\textwidth}{0.2pt}

%=================== Text begin here =============================================

\begin{center}
\LARGE\bf Controlling the transition between the Turing and antispiral patterns by using time-delayed-feedback$^{*}$   %% 论文题目
\end{center}

\footnotetext{\hspace*{-.45cm}\footnotesize $^*$Project supported by the National Natural Science Foundation of China with Grants No. 10975043, 10947166, the Natural Science Foundation of Hebei Province, China, with Grants No. A2011201006, A2010000185 and the Science Foundation of Hebei University.}
\footnotetext{\hspace*{-.45cm}\footnotesize $^\dag$Corresponding author. E-mail: Donglf@hbu.edu.cn}

\begin{center}
\rm He Ya-Feng, \ \ Liu Fu-Cheng,\ \ Fan Wei-Li, \ and  \ Dong Li-Fang$^{\dagger}$
\end{center}

\begin{center}
\begin{footnotesize} \sl
Hebei Key Laboratory of Optic-electronic Information Materials, College of Physics Science and Technology, Hebei University, Baoding 071002, China %%%% 地址 a)

%%% 更多地址依次往下延续
\end{footnotesize}
\end{center}

\begin{center}
%\footnotesize (Received X XX XXXX; revised manuscript received X XX XXXX)
          %% (Received 日 月 年; revised manuscript received 日 月 年)
\end{center}

\vspace*{2mm}

\begin{center}
\begin{minipage}{15.5cm}
\parindent 20pt\footnotesize
\indent The controllable transition between the Turing and antispiral patterns is studied by using time-delayed-feedback strategy in a FitzHugh-Nagumo model.
We treat the time delay as perturbation and analyze the effect of the time delay on the Turing and Hopf instabilities near the Turing-Hopf codimension-two phase space.
Numerical simulations show the transition between the Turing patterns (hexagon, stripe, and honeycomb), the dual-mode antispiral, and the antispiral by applying appropriate feedback parameters. The dual-mode antispiral pattern originates from the competition between the Turing and Hopf instabilities.
Our results have shown the flexibility of the time delay on controlling the pattern formations near the Turing-Hopf codimension-two phase space.
%%%% 论文摘要
\end{minipage}
\end{center}

\begin{center}
\begin{minipage}{15.5cm}
\begin{minipage}[t]{2.3cm}{\bf Keywords: }\end{minipage}
\begin{minipage}[t]{13.1cm}
pattern formation, Turing-Hopf bifurcations, time delay%%%%% 关键词
\end{minipage}\par\vglue8pt
{\bf PACS: 47.54.-r, 82.40.Ck, 82.40.Bj}
%%% PACS 分类码
%% 查询网址：http://www.aip.org/pacs
\end{minipage}
\end{center}

\section{Introduction}  %%% 节标题 1
\indent Spatiotemporal pattern formation has been extensively investigated in a variety of chemical, biological, and physical systems.$^{[1]}$ Since the observation of the Turing pattern and the spiral wave pattern in chlorite-iodide-malonic-acid$^{[2]}$ and Belousov-Zhabotinsky (BZ)$^{[3]}$ reaction, respectively, chemical reaction systems have attracted much attention on studying the pattern formation. In general, chemical systems can exhibit three types of properties: excitable, bistable, and Turing-Hopf. Some of these chemical reactions are light sensitive, such as the $Ru(bpy)_{3}^{2+}$ catalyst BZ reaction. The light-sensitive feature of the media makes it possible to control the spatiotemporal patterns by using certain strategy.$^{[4]}$ The control strategy can be classified to either external action such as the periodical forcing,$^{[5-11]}$ or internal one such as the time-delayed-feedback.$^{[12-16]}$ The time-delayed-feedback has been used widely due to its adaptive feature. For example, in excitable system, the rigid rotation of spiral can be stabilized by changing the domain diameter of feedback control.$^{[13]}$ In bistable system, time-delayed-feedback can control of the nonequilibrium Ising-Bloch bifurcation, which realizes the transformation between the spiral and labyrinth patterns.$^{[14]}$

\indent The systems of Turing-Hopf type have shown interesting spatiotemporal patterns, such as the hexagon, stripe, spiral and antispiral patterns. Some patterns are of desirable, such as the Turing hexagon grown in polystyrene film.$^{[17]}$ While some patterns are of avoid. Many methods have been attempted to control the Turing and Hopf patterns. For example, time-delayed-feedback with appropriate intensity has been used to suppress or induce the Turing patterns.$^{[18]}$ Recently, the systems near the Turing-Hopf codimension-two phase space have shown many fascinating patterns, such as the oscillatory Turing patterns obtained in a Brusselator model.$^{[19-21]}$ It is necessary to investigate the control of the spatiotemporal patterns near the Turing-Hopf codimension-two phase space. In this paper, we study the control of the transition between the Turing and antispiral patterns by using time-delayed-feedback strategy in a FitzHugh-Nagumo model. The effects of the time delay on the Turing and Hopf modes are analyzed by treating the time delay as perturbation. Numerical simulations show the flexibility of the time-delayed-feedback on controlling the transition between the Turing and antispiral patterns. We observe several dual-mode antispiral patterns and discuss their origins.

\section{Model}  %%% 节标题 2

\indent This work is based on a delayed FitzHugh-Nagumo model:
\begin{eqnarray}
  u_t &=& u -u^3 - v + D_{u} \nabla^2 u +F,\\
  v_t &=& \varepsilon(u-a_{1}v-a_{0})+D_{v} \nabla^2 v +G,
\end{eqnarray}
where the time delay is applied with the forms:
\begin{eqnarray}
  F=g_{u}(u(t-\tau)-u(t)),\\
  G=g_{v}(v(t-\tau)-v(t)),
\end{eqnarray}
here, variables $u$ and $v$ represent the concentrations of the activator and inhibitor, and $D_{u}$ and $D_{v}$ denotes their diffusion coefficients, respectively. The small value $\varepsilon$ characterizes the time scales of the two variables. The system described by Eqs.(1) and (2) can be either of Turing-Hopf, excitable, or bistable type. In this paper the parameter $a_{1}$ is chosen such that the system is Turing-Hopf type. The parameter $a_{0}$ determines the position of the uniform steady states on the nullclines of Eqs. (1) and (2). Unless otherwise specified, we choose $a_{0}$$=$$0$ such that the uniform steady states are $u_{0}$$=$$v_{0}$$=$$0$. The parameters $g_{u}$ and $g_{v}$ are the feedback intensities of variables $u$ and $v$, respectively. $\tau$ is the delayed time. In order to study the effect of the time delay on the patterns of Turing-Hopf type, we first execute the linear stability analysis. In the analysis, we treat the delay as a perturbation by expanding the feedback terms Eqs. (3) and (4) as:
\begin{eqnarray}
  u(t-\tau) &=& u(t) -\tau \frac{\partial u(t)}{\partial t},  \\
  v(t-\tau) &=& v(t) -\tau \frac{\partial v(t)}{\partial t}.
\end{eqnarray}
So, we obtain:
\begin{eqnarray}
  (1+\tau g_{u})u_t &=& u -u^3 - v + D_{u} \nabla^2 u,\\
  (1+\tau g_{v})v_t &=& \varepsilon(u-a_{1}v-a_{0})+D_{v} \nabla^2 v.
\end{eqnarray}
It can be seen that the terms of time is rescaled and determined by the delayed time and the feedback intensities. In the analysis, we consider the parameter $a_{0}$$=$$0$. So uniform steady state of the system reads: $u_{0}$$=$$v_{0}$$=$$0$. Then, we perturb the uniform steady state with small spatiotemporal perturbation ($\delta$$u$, $\delta$$v$) $\sim$ $exp(\lambda t+ikr)$, and obtain the following matrix equation for eigenvalues:
\begin{displaymath}
\mathbf{ }
\left( \begin{array}{cc}
1 & -1  \\
\varepsilon & -\varepsilon a_{1}
\end{array} \right)
\left( \begin{array}{cc}
\delta u \\
\delta v
\end{array} \right)=0,
\end{displaymath}
So we get the following quadratic equation for the eigenvalues:
\begin{equation}
A \lambda^2-B\lambda+C = 0,
\end{equation}
where,
\begin{equation}
A = (1+\tau g_{u})(1+\tau g_{v}),
\end{equation}
\begin{equation}
B = (1+\tau g_{u})(\varepsilon a_{1} + k^2 D_{v})-(1 + \tau g_{v})(1 - k^2 D_{u}),
\end{equation}
\begin{equation}
C = \varepsilon - (1 - k^2 D_{u})(\varepsilon a_{1} + k^2 D_{v}).
\end{equation}
The dispersion relations $\lambda(k)$ of the system are defined by the roots of Eq. (9),
 \begin{equation}
\lambda_{1,2} = \frac{B \pm \sqrt{B^2 -4AC}}{2A}.
\end{equation}
 \indent The positive real parts of $\lambda(k)$ give rise to Hopf instability and Turing instability, which occur at $k$$=$$0$ and $k$$\neq$$0$, respectively. In the absent of time delay, the stability diagram of the uniform steady state in the ($a_{1}$, $\varepsilon$) plane is shows in Fig.1. The Turing bifurcation line (solid line) and the Hopf bifurcation line (dash line) cross at one point. Applying time delay would affect the two bifurcation lines. Here, we take the parameters set as ($a_{1}$, $\varepsilon$)=($0.5$, $2$) as indicated by the open circle in Fig. 1, which is beyond the Turing bifurcation line and locates on the Hopf bifurcation line. Our emphasis is on controlling the pattern formation by using the strategy of time-delayed-feedback near the Turing-Hopf codimension-two phase space. In the following we will discuss the effect of time delay on the Hopf and Turing instabilities, respectively. Our analysis shows that the time delay affects both the Hopf and the Turing instability.

\begin{figure}[htbp]
  \begin{center}\includegraphics[width=8cm,height=6cm]{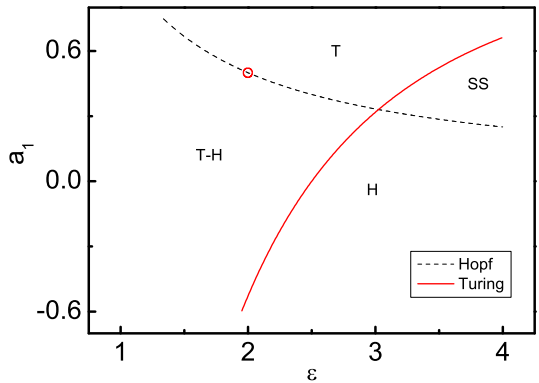}
  \caption{Bifurcation diagram of the model system. Other parameters are: $D_{u}$$=$$0.01$, $D_{v}$$=$$0.1$, $a_{0}$$=$$0.0$, $\tau$$=$$0.0$, $g_{u}$$=$$g_{v}$$=$$0.0$.}
  \end{center}
\end{figure}

\indent The threshold condition of Hopf instability reduced from the linear stability analysis can be expressed as
\begin{equation}
a_{1}^{H} = \frac{(1 + \tau g_{v})}{\varepsilon (1 + \tau g_{u})},
\end{equation}
  and the corresponding frequency of oscillation is
 \begin{equation}
\omega_{0} = \sqrt{\frac{\varepsilon}{(1 + \tau g_{u})(1 + \tau g_{v})}-\frac{1}{4}\Big[\frac{\varepsilon a_{1}}{1 + \tau g_{v}}+\frac{1}{1 + \tau g_{u}}\Big]^{2}}.
\end{equation}

\indent Eqs. (14) and (15) show that the strategy of time-delayed-feedback plays an important role on controlling the bifurcation line and the oscillatory frequency of Hopf instability. It can be seen from Eq.(14) that applying positive feedback to variable $u$ (i.e., $g_{u}$$>$$0$) and/or negative feedback to variable $v$ (i.e., $g_{v}$$<$$0$) is functionally equivalent to increasing $\varepsilon$ and/or decreasing $a_{1}$, and vise versa. We want to mention that in a special case of $g_{u}$$=$$g_{v}$, the feedback does not affect the bifurcation line of Hopf instability any more. However, it still changes the oscillatory frequency of Hopf mode as is illustrated in Eq. (15). Fig. 2 (c) shows the dependence of the oscillatory frequency on the feedback intensities when applying the feedback with identical intensities $g_{u}$$=$$g_{v}$. The oscillatory frequency decreases dramatically with feedback intensities. The stronger the intensities of the negative feedback are, the faster the Hopf mode oscillates. In the case of $g_{u}$$\neq$$g_{v}$, the dependence of the oscillatory frequency on the feedback intensity $g_{u}$ and $g_{v}$ are shown in Fig. 2 (a) and (b), respectively. The solid line part in Fig. 2 represents that the system is beyond the bifurcation line of Hopf instability. When the feedback $g_{u}$ is applied individually, the oscillatory frequency increases from zero and then decreases with $g_{u}$ as indicated by the solid line in Fig. 2 (a). When the feedback $g_{v}$ is applied individually, the oscillatory frequency decreases with $g_{v}$ as shown in Fig. 2 (b). Therefore, the time-delayed-feedback provides a way to change the oscillatory frequency of Hopf mode even under the condition that the intensity of Hopf mode remains unchanged.

\begin{figure}[htbp]
  \begin{center}\includegraphics[width=8cm,height=12cm]{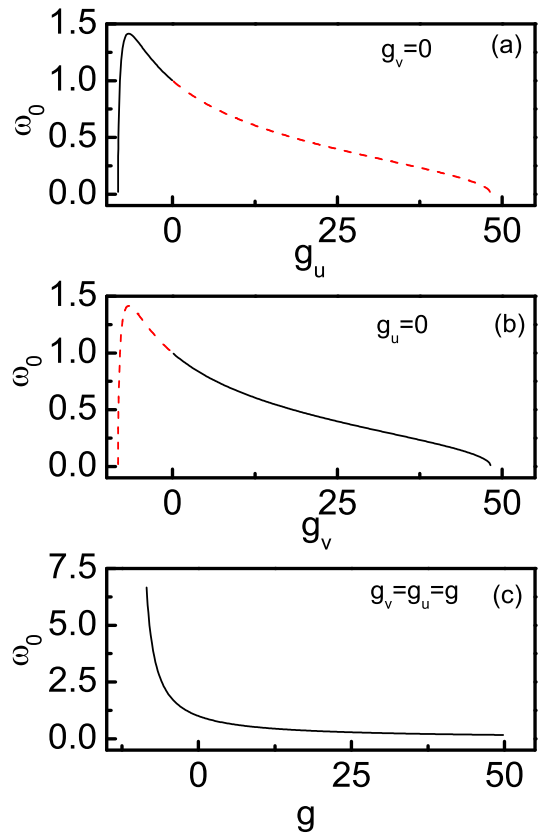}
  \caption{Dependence of the oscillatory frequency of Hopf mode on the feedback intensities in three cases: (a), $g_{v}$$=$$0$; (b), $g_{u}$$=$$0$; (c), $g_{u}$$=$$g_{v}$$=$$g$. Other parameters are: $D_{u}$$=$$0.01$, $D_{v}$$=$$0.1$, $a_{1}$$=$$0.5$, $a_{0}$$=$$0.0$, $\varepsilon$$=$$2.0$, $\tau$$=$$0.1$.}
  \end{center}
\end{figure}

\indent Fig. 3 shows the dependence of the wave vector of the most unstable Turing mode on the feedback intensities. It can be seen that when the feedback $g_{u}$ ($g_{v}$) is individually applied to the system the wave vector increases (decreases) with the feedback intensity. This reveals that the wavelength of the Turing patterns can be controlled by using appropriate time-delayed-feedback. In a special case of $g_{u}$$=$$g_{v}$, the time delay do not affect the wave vector of the Turing mode any more as indicated by the dash line in Fig. 3. In this case, the time delay is equivalent to rescaling time as shown in Eqs. (7) and (8) and keeping the ratio of diffusion coefficients unchanged. Thus the wavelength of Turing pattern is unaffected.

\begin{figure}[htbp]
  \begin{center}\includegraphics[width=8cm,height=6cm]{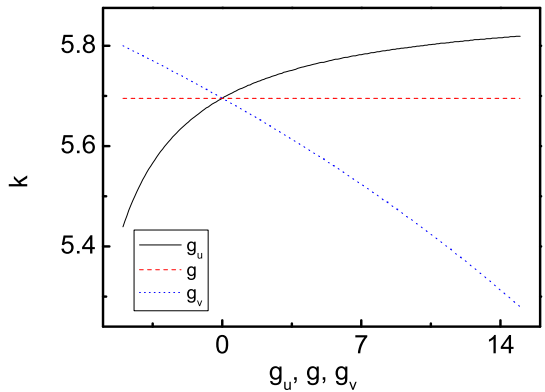}
  \caption{Dependence of the wave vector of the most unstable Turing mode on the feedback intensities in three cases: solid line, $g_{v}$$=$$0$; dash line, $g_{u}$$=$$g_{v}$$=$$g$; dot line, $g_{u}$$=$$0$. Other parameters are: $D_{u}$$=$$0.01$, $D_{v}$$=$$0.1$, $a_{1}$$=$$0.5$, $a_{0}$$=$$0.0$, $\varepsilon$$=$$2.0$, $\tau$$=$$0.1$.}
  \end{center}
\end{figure}

\section{Two-dimensional numerical simulation}

\indent In order to illustrate the above analytical results on controlling the pattern formation by using the strategy of time-delayed-feedback, we have carried out numerical simulation in two dimensions. The grid sizes are $200$$\times$$200$ which represents domain sizes $20$$\times$$20$ s.u., and the time step is $\triangle$$t$$=$$0.01$ t.u.. The boundary condition is zero flux boundary. In the above analysis, the time delay was treated as a perturbation, therefore the delayed time $\tau$ should be a small value. The feedback intensities $g_{u}$ and $g_{v}$ can be of any values. Eqs. (7) and (8) have shown that the delayed time $\tau$ and the feedback intensity $g_{u}$ ($g_{v}$) are incorporated together, which means that the effect of small delayed time and large feedback intensity, for example ($\tau$$=$$0.1$, $g_{u}$$=$$10.0$), is equivalent to that of large delayed time and small feedback intensity ($\tau$$=$$10.0$, $g_{u}$$=$$0.1$). In the following we only show the numerical results with small delayed time $\tau$$=$$0.1$.

\begin{figure}[htbp]
  \begin{center}\includegraphics[width=8cm,height=6cm]{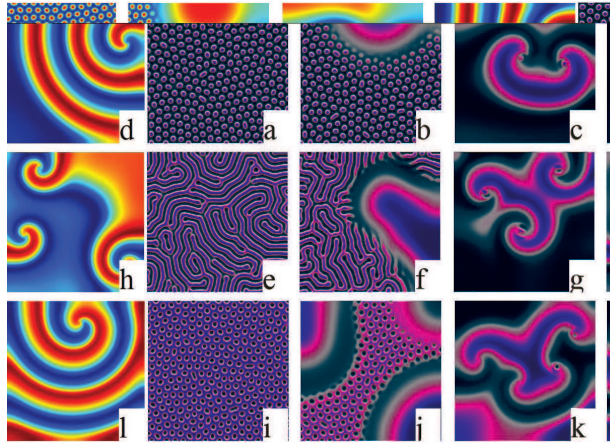}
  \caption{Transition from Turing to antispiral patterns with increasing $g_{v}$ in three cases: $a_{0}$$=$$-0.1$, (a)-(d); $a_{0}$$=$$0.0$, (e)-(h); $a_{0}$$=$$0.1$, (i)-(l). The feedback intensities $g_{v}$ in (a)-(d), (e)-(h),(i)-(l) are $0.0$, $5.0$, $6.0$, $14.0$, respectively. Other parameters are: $D_{u}$$=$$0.01$, $D_{v}$$=$$0.1$, $a_{1}$$=$$0.5$, $\varepsilon$$=$$2.0$, $\tau$$=$$0.1$, $g_{u}$$=$$0.0$.}
  \end{center}
\end{figure}

\indent Fig. 4 shows three cases for controlling the transition between the Turing and antispiral patterns when the feedback $g_{v}$ is individually applied to the system. In the case of $a_{0}$$=$$-0.1$, Fig. 4 (a)-(d) show the transition from the hexagon to antispiral patterns with increasing $g_{v}$. When $g_{v}$$=$$0.0$, it is a hexagon pattern as shown in Fig. 4 (a). This is because that the Turing mode is positive while the Hopf mode is negative as illustrated in Fig. 5. With increasing the feedback $g_{v}$, the Hopf mode becomes stronger and stronger as shown in Fig. 5, which induces oscillating patterns such as the antispiral pattern. The Turing mode and the Hopf mode will compete, which results in the coexistence of the stationary Turing pattern and the oscillating patterns as shown in Fig. 4 (b). If the Hopf mode is dominant one the oscillating pattern swallows the stationary Turing pattern gradually, and if the Turing mode is prominent the stationary Turing pattern will occupy the whole domain finally.

\begin{figure}[htbp]
  \begin{center}\includegraphics[width=8cm,height=6cm]{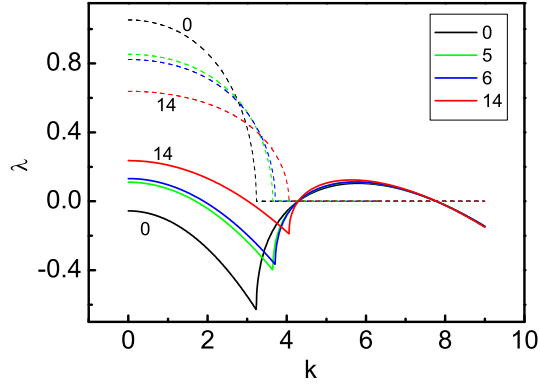}
  \caption{Dispersion relation of the delayed model in four case: $g_{v}$$=$$0$, $5$, $6$, $14$. The solid (dash) lines represent the real (imaginary) part of the eigenvalue. The numbers indicated in the label represent the values of $g_{v}$. Other parameters are: $D_{u}$$=$$0.01$, $D_{v}$$=$$0.1$, $a_{1}$$=$$0.5$, $a_{0}$$=$$-0.1$, $\varepsilon$$=$$2.0$, $\tau$$=$$0.1$, $g_{u}$$=$$0.0$.}
  \end{center}
\end{figure}

\indent Fig. 4 (c) shows a dual-mode antispiral pattern which exhibits the competition between the Hopf and Turing modes near the Turing-Hopf codimension-two phase space. The antispiral pattern originates from the negative dispersion $d\omega/d|k|$$<$$0$ and the faster bulk oscillation $\omega_{0}$$>$$\omega_{k}$ as indicated by the dash lines in Fig. 5. Here, we focus on the core region of the antispiral pattern. As is well known, the tip of oscillating antispiral is a topological defect, and the amplitude near the tip is zero. However, in the present case of competition between the Hopf and Turing instabilities, the amplitude near the tip is not zero yet. The core of antispiral pattern contains stationary hexagonal spots originating from the Turing instability as shown in Fig. 4(c). The wavelength of the hexagon pattern near the core is about $1.1$ s.u. which corresponds to the wave vector of the most unstable Turing mode $k$$=$$5.7$. The amplitude of the hexagon pattern is lower than that of the antispiral pattern as shown by the surface plot in Fig. 6. Far away from the core region, the hexagon pattern is damped by the Hopf oscillation. This is similar to the dual-mode spirals observed by Mau$^{[22]}$ and Kepper$^{[23]}$, respectively. With increasing $g_{v}$, the amplitude of the antispiral pattern increases gradually. The hexagon spots near the core is suppressed by the oscillation of large amplitude. The core of the antispiral pattern returns to a normal one as shown in Fig. 4 (d).

\begin{figure}[htbp]
  \begin{center}\includegraphics[width=8cm,height=6cm]{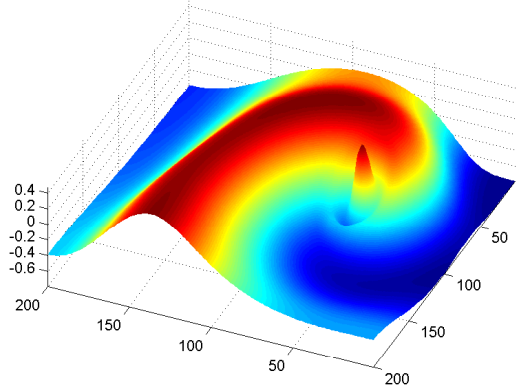}
  \caption{Surface plot of the core of a dual-mode antispiral pattern. The amplitude of the hexagonal spots near the core is lower than that of the antispiral pattern. The parameters are the same as that in Fig. 4 (c).}
  \end{center}
\end{figure}

\indent In the case of $a_{0}$$=$$0.0$, Fig. 4 (e)-(h) show the transition from the stripe to antispiral patterns with increasing $g_{v}$. This progress is similar to the aforesaid transition. The core of this dual-mode antispiral contains stripe pattern as shown in Fig. 4 (g). In the case of $a_{0}$$=$$0.1$, Fig. 4 (i)-(l) show the transition from the honeycomb to antispiral patterns with increasing $g_{v}$. The core of this dual-mode antispiral contains honeycomb pattern as shown in Fig. 4 (k). In a word, we can control the transition between the Turing and antispiral patterns by using time-delayed-feedback.

\indent The dual-mode antispiral is similar to the observation by Yuan \textit{et al} in a CIMA reaction.$^{[24]}$ They called it Turing-Hopf mixed state and attributed its formation to the 3D effect of the reaction medium, where the Turing pattern and the antispiral pattern occur in different places in the third dimension. However, in our case the dual-mode antispiral pattern occurs during the transition as increasing $g_{v}$. In this progress the Hopf mode becomes stronger and stronger, and the amplitude of the antispiral pattern increases gradually. When the amplitudes of the Turing pattern and the antispiral pattern are comparable as indicated in Fig. 5, the Turing pattern sustains near the core of the antispiral pattern, which forms the dual-mode antispiral pattern as shown in Fig. 4 (c), (g) and (k). Far away from the core region the Turing pattern is damped by the Hopf oscillation. With increasing $g_{v}$ continuously, the Hopf mode becomes strong enough. The amplitude of the antispiral pattern is much larger than that of the Turing pattern. The Turing patterns are damped nearly in the whole domain. Thus we observe the normal antispiral patterns as shown in Fig. 4 (d), (h) and (l).

\indent In the above simulations, we have shown the controllable transition between the Turing patterns and the antispiral pattern by increasing the feedback intensity $g_{v}$. As illustrated in Eq. (14), the effect of increasing $g_{v}$ is equivalent to decreasing $g_{u}$. The aforesaid transition from the Turing patterns (hexagon, stripe, and honeycomb) to the antispiral pattern can be reproduced by decreasing the feedback $g_{u}$. This has been confirmed by extensive numerical simulations (not shown here). Therefore, the time-delayed-feedback provides a flexible strategy to control the Turing and antispiral patterns near the Turing-Hopf codimension-two phase space.

\section{Conclusions}

\indent We have studied the transition between the Turing and antispiral patterns controlled by time-delayed-feedback in a delayed FitzHugh-Nagumo model. We treated the time delay as perturbation and analyzed the effect of the time delay on the Turing and Hopf instabilities near the Turing-Hopf codimension-two phase space. The time delay affects the wave vector of the Turing mode and the oscillation frequency of the Hopf mode. Analyzes show that the effect of small delayed time and large feedback intensity on the system is equivalent to that of large delayed time and small feedback intensity. The time-delayed-feedback provides a flexible way to control the pattern formation. The transition between the Turing patterns (hexagon, stripe, and honeycomb), the dual-mode antispiral, and the antispiral is numerically studied by applying appropriate feedback parameters. The dual-mode antispiral patterns originate from the competition between the Turing and Hopf modes. Our results could contribute to the control of pattern formations in the light-sensitive BZ or chlorine dioxide-iodine-malonic-acid chemical reactions.

%%%% 参考文献排版格式：

\end{CJK*}  %% 结束中文、日文、韩文使用环境
\end{document}